# A NOVEL UNIVERSAL PHOTOVOLTAIC ENERGY PREDICTOR

**Nirupam Bidikar[1], Kotoju Rajitha[2], P. Usha Supriya[3]**

[1]Student, Department of Computer Science, Mahatma Gandhi Institute of Technology, Hyderabad, Telengana, India.

[2]Assistant Professor, Department of Computer Science, Mahatma Gandhi Institute of Technology, Hyderabad, Telengana, India.

[3]Student, Department of Computer Science, Mahatma Gandhi Institute of Technology, Hyderabad, Telengana, India.

## ABSTRACT

Solar energy is one of the most economical and clean sustainable energy sources on the planet. However, the solar energy throughput is highly unpredictable due to its dependency on a plethora of conditions including weather, seasons, and other ecological/environmental conditions. Thus, the solar energy prediction is an inevitable necessity to optimize solar energy and also to improve the efficiency of solar energy systems. Conventionally, the optimization of the solar energy is undertaken by subject matter experts using their domain knowledge; although it is impractical for even the experts to tune the solar systems on a continuous basis. We strongly believe that the power of machine learning can be harnessed to better optimize the solar energy production by learning the correlation between various conditions and solar energy production from historical data which is typically readily available. For this use, this paper predicts the daily total energy generation of an installed solar program using the Naïve Bayes' classifier. In the forecast procedure, one-year historical dataset including daily moderate temperatures, daily total sunshine duration, daily total global solar radiation and daily total photovoltaic energy generation parameters are used as the categorical-valued features. By implementing this approach, we observe a noticeable improvement in the accuracy and sensitivity and also explore the how photovoltaic energy generation is affected by various solar parameters.

**Index Terms:** Solar energy, Photovoltaic Energy, Machine Learning, Naive Bayes' classifier

## I. INTRODUCTION

With the advancement in technology and the growing population, energy production has to be increased to meet the growing demands. It is estimated that Electricity production on our planet will grow to an alarming 36.5 trillion kWh by the year 2040. Our reliance on resources like gas, coal, and natural gas will soon come to an end, as we are becoming increasingly aware that these resources are being depleted at a dangerously unsustainable rate. Consequently, there is increasing focus on finding alternative *sustainable* sources of energy. For example, international yearly investments in renewable energy resources are significantly increasing and solar energy has become one of the most preferred energy sources since it is one of the most renewable sources of energy supply. In fact, while total solar energy production is rising at an average of 8.3% [1,2] each year and is predicted to rise to 15.7% during 2012 through 2040 [3,4].

While the solar energy is sustainable, the source of energy is itself somewhat dependent on various environmental and ecological factors. For operations, planning, and scheduling purposes, it is critical to find correlation between the various concomitant historical data and the solar energy availability. Further, it would be useful to learn these correlations automatically through machine learning algorithms so that it can be incorporated into enterprise workflows for optimizing the energy utilization operations. Specifically, our motivation for this research is to explore methods of predicting the solar energy levels from available historic data.

The research focus of this paper, a universal photovoltaic energy predictor, is an effective system to forecast the radiation later on in a bid to cut down the use of non-renewable sources of energy and taking a step forwards towards a green and sustainable future. Utilizing a Naive Bayes classifier, we shall have the ability to forecast the categorical worth of Photovoltaic Energy which could be generated at a certain location. It might serve another function which is to determine whether that particular place would be suitable to establish a new solar energy plant project and whether the surrounding regions could be powered solely using the renewable energy generated from the newly established plant.

# II. LITERATURE SURVEY

Prediction of the solar energy levels is not a new problem. A number of researchers have attempted various methods to approach this problem.

Creayla et al. [5] Ibrahim et al. [6] leveraged random woods, artificial neural networks, firefly algorithm-based random woods and firefly algorithm-based artificial neural networks for global solar radiation prediction. Their average bias errors were calculated as 3.6146 percent, 6.9889 percent, 2.8631percent and 4.9725 percent, respectively. Wang et al. [7] used multilayer perceptron, generalized regression neural network and radial basis neural network for global solar radiation. Errors of multilayer perceptron, generalized regression neural network and radial basis neural system models were acquired as 1.61 MJ/m2, 1.56 MJ/m2 and also 1.68 MJ/m2, respectively. Belaid et al. [8] used the support vector machines for the forecast of global solar radiation. Belaid et. Al's support vector machine calculated the normalized root square mistake to 7.442 percent. Wu et. Al. used self-respecting map-based optimally pruned intense learning system, back-propagation neural system and autoregressive integrated moving average model for solar power calling [9].

Jovic et al. [10] proposed the adaptive neuro-fuzzy inference method for solar power prediction using relative humidity, mean sea level, dry-bulb fever and wet-bulb temperature parameters. Even the dry-bulb temperature and relative humidity points are identified as the ones that were effective. Hassan et al. [11] utilised from support vector machines, adaptive neuro-fuzzy inference method and multilayer perceptron for international solar radiation prediction. The coefficients of conclusion of the day number- and temperature-based versions calculated greater than 85%. Hassan et al. [12] suggested the sine wave version, Lorentzin correlation model, Gaussian correlation model, cosine wave model, polynomial model and hybrid sine and cosine wave correlation model to gauge the international solar radiation models. Yesilbudak et. Al. [13] clustered the multipurpose monthly average insolation period data employing the K-Means algorithm and discovered the most successful cities in terms of the solar energy penetration

## Disadvantages of Previous Systems

All these preceding implementations lacked the capability to dynamically alter the place in which the solar energy would be to be predicted along with the capacity to change the period of time of their datasets were employed to train the classifier. Additionally, conversion of raw data in categorical format is time consuming.

The authors of the paper **A Novel Application of Naive Bayes Classifier in Photovoltaic Energy Prediction,** Ramazan Bayindir, Mehmet Yesilbudak, Medine Colak, Naci Genc [14] used data mining approach to predict the photovoltaic energy. Data mining[15] is a knowledge discovery procedure in databases and it's based upon the sorting, searching, outlining and analyzing massive amounts of information in the database. There are numerous data mining techniques utilized in the literature including Artificial Neural Networks, Support Vector Machines, including Naïve Bayes classifier, K-Nearest

Neighbor algorithm and genetic algorithm, etc... Five distinct tags namely "Very Low", "Low", "Medium", "High", and "Very High" are utilized with different selection and parameters. Along with such tests, the supply of daily average temperatures, daily complete sunshine duration, daily total global solar radiation and daily complete photovoltaic energy production values will also be known as for the energy forecast. The precision was within acceptable limits but their experimental work focused on only one experimental setup with features comprising on very limited sensor data. We felt that their work did not address the need for a majority of users from different geographical regions. Further they did not leverage additional sensor data that is available from public domain sources such as NASA. Finally, there are data volume issues related to taking specific sensors and refactoring them to specific format required for machine learning algorithms. To handle these requirements, we have proposed a novel Naive Bayes classifier-based system[16] and trained it using parameters – Average Temperature, Clearness index, and Incident Solar Radiation having a season's worth of information of a specific place which is topic of the next section below.

# III. PROPOSED SYSTEM

In summary, while a number of researchers have focused on the solar energy prediction, we felt that a majority of effort was directed narrowly towards predicting solar energy at a specific location. More recently, we found research [14] which was directly making the prediction agnostic to the place of the solar plant. However, the effort was directed at incorporating sensors into the solar plant and the experiments were conducted on a specific set up at only one location. In order to make solar energy prediction agnostic to place, we found it desirable to include sensing location as a parameter in our experimental work. The idea is beguilingly simple yet very effective: We illustrate a Machine Learning method that chooses the coordinates of a place as input by the end-user and by creating an API call with NASA's POWER[17], we recover the parameters essential to construct a data sample that is relevant to the end-user geographic location. Another advantage of our method is that it leverages a multitude of the relevant feature representation that is freely available at NASA weather observatory. The dataset is curated for lost values and categorical tags are attached to every value (very low, low, moderate, high and very high). The resulting representation is fed into the Naive Bayes' classifier. As noted below in the experimental results section, our experimental methodology splits the dataset into training and test sets. The training utilizes the training samples while the performance (e.g., precision) is measured with the test collection. In addition, to further validate generalization power of our models, we create another dataset for the subsequent year to further examine the validity of this classifier. In a typical user scenario, the predictions are likely done by entering the values of these parameters (average temperature, clearness index and incident radiation) and the output is a categorical tag to which the energy value will likely belong to.

### Advantages of Proposed System

The outcomes of the experiment demonstrate a greater accuracy in a selection of 90-96 percent. We highlighted a simple method that takes solar plant location values into account and thus making the system more useful almost anywhere globally. As a result, our method not only conserves time, but also saves the expense of additional sensors needed. Thanks to NASA's feature representation, our method relies on very few features which in turn results in decreased amount of attributes passed to the classifier, improving training time and convergence. We also note that our approach has resulted in better outcomes compared to other implementations. Unlike current solutions, we could forecast the precision for any place.

# IV. IMPLEMENTATION

In this section we described the details of our implementation and experimental methodology.

The entire system contains 3 core components.

- User Interface
- Dataset
- Classifier

## User Interface

A simple user interface is intended to permit the user to input the essential parameters and request their forecasts. The inputs for predictions would be the parameters "T_avg", "KT" and also "S_mod" and the outcome would be the title of this tag where the PVE for this information would appeal to. The UI was built with the Tkinter library that comes pre-installed with python and does not require any extra dependencies. It's a lightweight framework which is helpful in developing small scale GUIs.

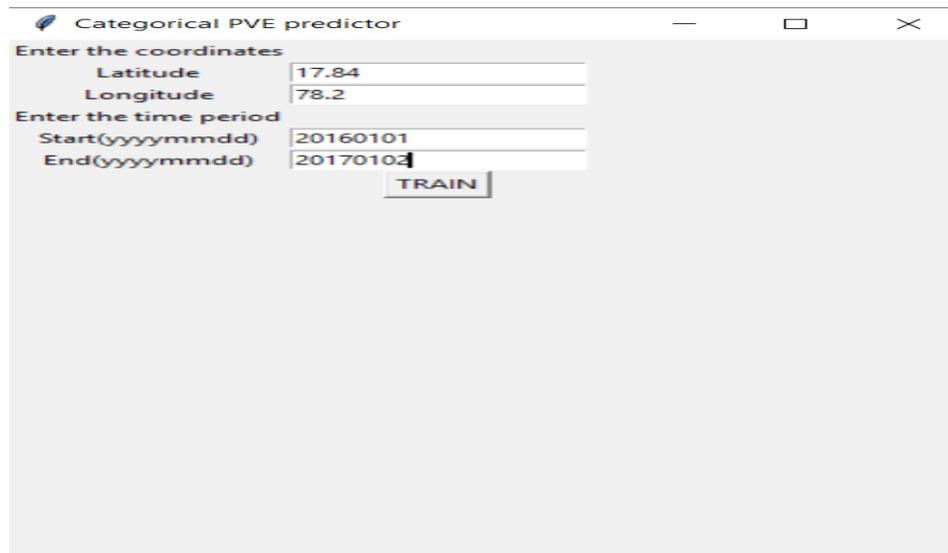

**Fig: 1 UI co-ordinates**

We utilize the fundamental elements - tags and text boxes to produce the basic form like construction for the consumer to input the coordinates of a place and also to opt for a time interval. Values in the input elements are routed to the next module that utilizes those to create a URL. The URL is subsequently utilized to create the API call to recover the exact information.

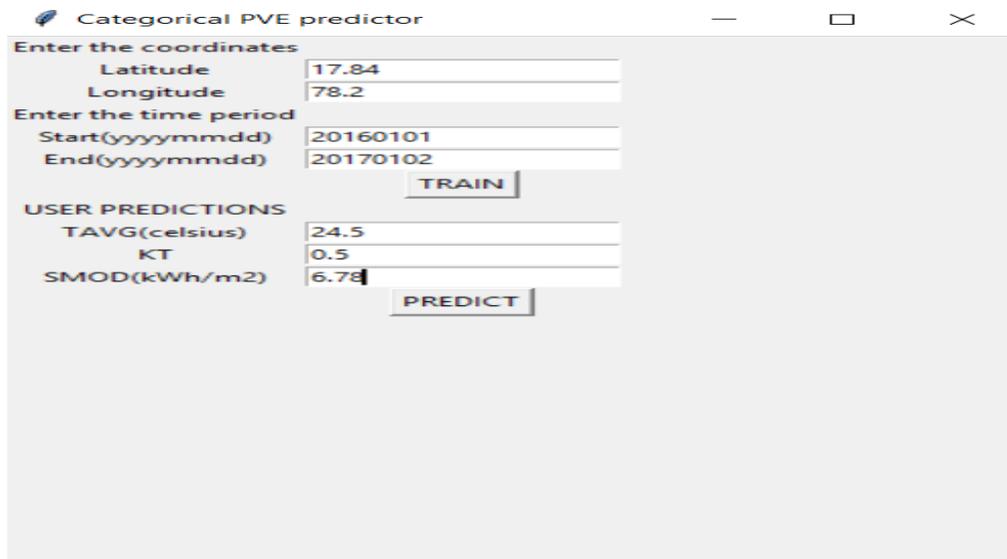

**Fig: 2 Predictor**

When the classifier is trained, then the User Prediction section shows up. The values to the attributes are typed here to the PVE will be predicted. The output generated is as follows.

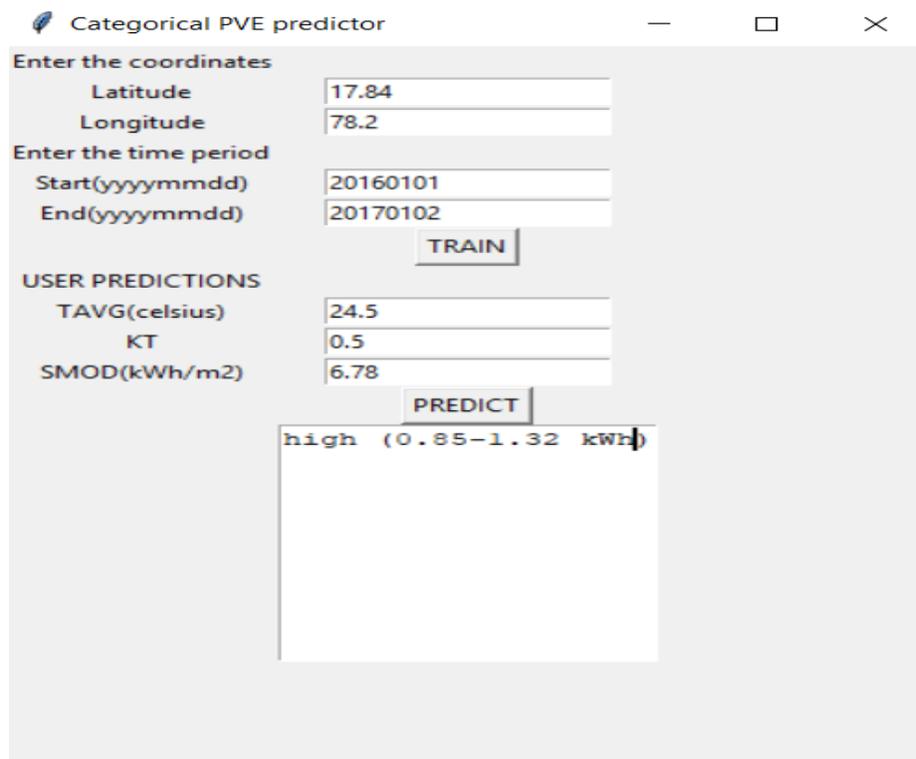

**Fig: 3 Result**

## Dataset

The dataset employed to train the classifier is generated utilizing the POWER API[17] offered by NASA. The API yields several parameters such as minimum and maximum temperatures, clearness index of the sky, incident solar radiation onto a solar panel. The input obtained from the user is used to construct the API call and a request is made to the server. Base URL for the API is given below.

https://power.larc.nasa.gov/cgi-bin/v1/DataAccess.py?&request=execute&identifier=SinglePoint¶meters=[features]&startDate=[begin]1&endDate=[end]&userCommunity=SSE&tempAverage=DAILY&outputList=CSV&lat=[latitude]&lon=[longitude]

This data was obtained from the NASA Langley Research Center (LaRC) POWER Project funded through the NASA Earth Science/Applied Science Program. The resulting output is a JSON tree of the values of these parameters requested within the time frame which was also provided by the user. A sample JSON tree is shown below (Figure 4). This data was created for the year 2016-17 for the coordinates 17.41,78.47.

```json
{
    "features": [
        {
            "geometry": {
                "coordinates": [
                    78.47861,
                    17.41381,
                    542.53
                ],
                "type": "Point"
            },
            "properties": {
                "parameter": {
                    "ALLSKY_SFC_SW_DWN": {
                        "20170101": 4.99,
                        "20170102": 4.95,
                        "20170103": 5.0,
                        "20170104": 5.09,
                        "20170105": 5.15,
                        "20170106": 5.15,
                        "20170107": 5.06,
                        "20170108": 5.0,
                        "20170109": 4.94,
                        "20170110": 4.68,
                        "20170111": 5.09,
                        "20170112": 5.17,
                        "20170113": 4.68,
                        "20170114": 4.94,
                        "20170115": 4.97,
                        "20170116": 5.21,
                        "20170117": 5.19,
                        "20170118": 5.26,
                        "20170119": 5.09,
                        "20170120": 5.26,
                        "20170121": 5.29,
                        "20170122": 5.2,
                        "20170123": 5.27,
                        "20170124": 5.34,
                        "20170125": 5.37,
                        "20170126": 3.52,
                        "20170127": 5.01,
                        "20170128": 5.32,
                        "20170129": 5.65,
```

**Figure 4: A Snippet of the JSON Tree**

This JSON response is subsequently converted into a DataFrame object using the Pandas library. Dataframe is an object which stores information in columns and rows. The DataFrame is subsequently examined for missing values and such records are eliminated. Depending on the information obtained, appropriate labels to every value of photovoltaic energy are given belonging in one of the five categories, i.e., "very low", "low", "moderate", "high", and "very high".

To further analyze the features and find correlation among them, we plotted graphs to show the distribution of features against time at the given location. The plots for distributions of Average Temperature, Clearness Index and Incident Solar Radiation against time are given below (Figure 5, 6, 7).

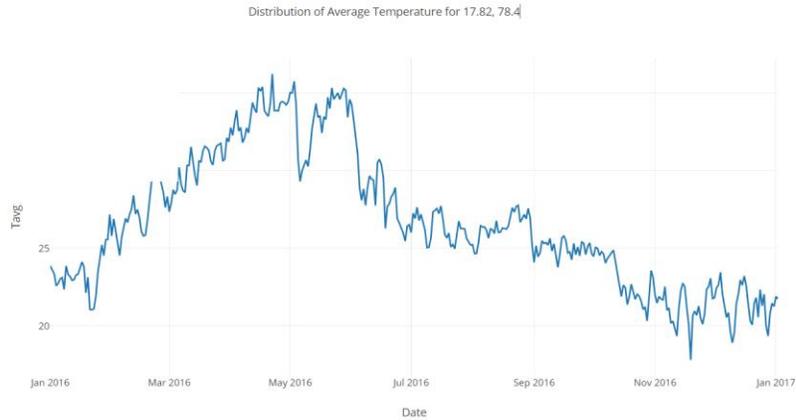

**Fig: 5 Distribution of Average Temperature**

The above graph shows the variations in average temperature for the year 2016 at the coordinates (longitude=17.82, latitude=78.47) for the year 2016.

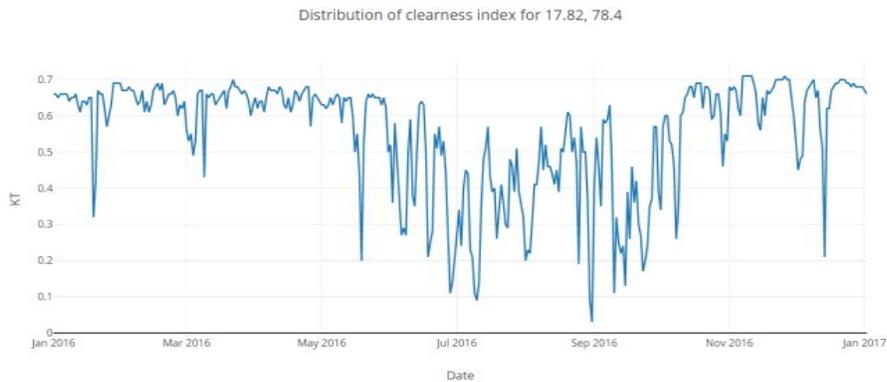

**Fig: 6 Distribution of Clearness Index**

The above graph shows the variations in clearness index at the coordinates (longitude=17.82, latitude=78.47) for the year 2016. Clearness index accounts for various factors like cloud coverage, dust, mist etc.

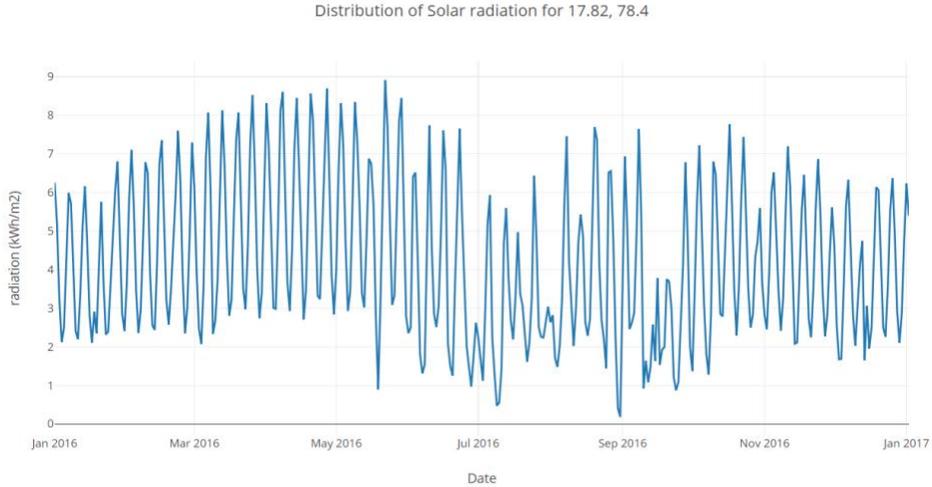

Fig: 7 Distribution of solar radiation

The above graph shows the variations in solar radiation received at the coordinates (longitude=17.82, latitude=78.47) for the year 2016. This is the solar energy incident on a horizontal surface.

We then make use of Pv-Lib[18] to find out the elevation angle for the required location at given time. This is used to calculate the perpendicular component of the radiation which strikes the solar panel.

### Classifier:

The classifier of choice has been Naive Bayes[16]. It's a collection of algorithms based on Bayes' Theorem that works on the principle of class conditional independence which translates to every feature being independent of the value of any other feature. The classifier makes statistical estimations by calculating the conditional probability. The Bayes' Theorem is explained as -

$$P(A|B) = (P(B|A) * P(A)) \div P(B)$$

Where P(A|B) is the probability of event A in case B occurs which is also known as the posterior probability, P(B|A) is the probability of event B in case A occurs, P(A) is the probability of event A occurring and P(B) is the probability of event B occurring.

Unlike the previous paper, we went ahead with the **Gaussian Naïve Bayes**[19] implementation. This variant of the classifier is an event model and can accept continuous valued features. The main assumption made in this approach is that the values associated with each class are distributed according to the Gaussian distribution.

Consider a continuous valued feature z from the dataset. For some associated class $C_k$, mean $\mu_k$ and variance $\sigma^2_k$ are calculated. Probability distribution for some observed value x can be computed as follows.

$$p(z = x | C_k) = \frac{1}{\sqrt{2\pi\sigma_k^2}} e^{-\frac{(x-\mu_k)^2}{2\sigma^2}}$$

We use the Scikit-Learn[20] library to instantiate and train the classifier. The dataset is split into training and test examples and the classifier is trained. Later, this trained classifier can be employed to predict the values of the next calendar year. The attributes selected to train the classifier are "T_avg" which is the average temperature of that location, "KT" that is the clearness index of the sky and "S_mod" which is the amount of solar radiation received at that place calculated at 12:40 PM local time. The test parameter accuracy is calculated as follows.

$$Accuracy = \frac{(TP + TN)}{(TP + TN + FP + FN)}$$

# V. RESULTS AND DISCUSSIONS

The accuracy achieved in the previous system was around 82.17%. We were able to attain an accuracy in the range of 90.607 – 96.124% with no categorical encoding of any dataset and with a split ratio of 0.35. The confusion matrix for this situation is given below (Table 1).

## Confusion Matrix:

A confusion matrix allows us to visualize the operation of an algorithm especially in the case of supervised learning. It provides a simplistic representation to highlight the difference between the actual values and the predicted values.

The confusion matrix (Table 1) was generated when the classifier was trained with the 2016-2017 dataset and was used to predict the values for the period 2017-18. Total number of predictions made were 365 with an accuracy of 94.2465%. The location chosen for this experiment is exactly the same as that of the previous paper, Van, Turkey with the input coordinates being 38.499, 43.365.

| (predicted) / (actual) | Very Low | Low | Moderate | high | Very high |
|---|---|---|---|---|---|
| **Very low** | 27 | 3 | 0 | 0 | 0 |
| **Low** | 0 | 70 | 6 | 0 | 0 |
| **Moderate** | 0 | 3 | 62 | 2 | 0 |
| **High** | 0 | 0 | 1 | 68 | 3 |
| **Very High** | 0 | 0 | 0 | 3 | 117 |

**Table 1: Confusion Matrix**

## Additional Results:

We selected 4 random latitudes (-60,-30,40,72) and longitudes (-150,-75,90,140) which generated 16 distinct locations and tested these with the classifier and finding out accuracies for the two time periods 2016-2017 (Figure 8) and 2017-2018 (Figure 9). The accuracy of 2016-2017 is based on the test set while, the 2017-2018 is a completely new dataset. The results are shown below.

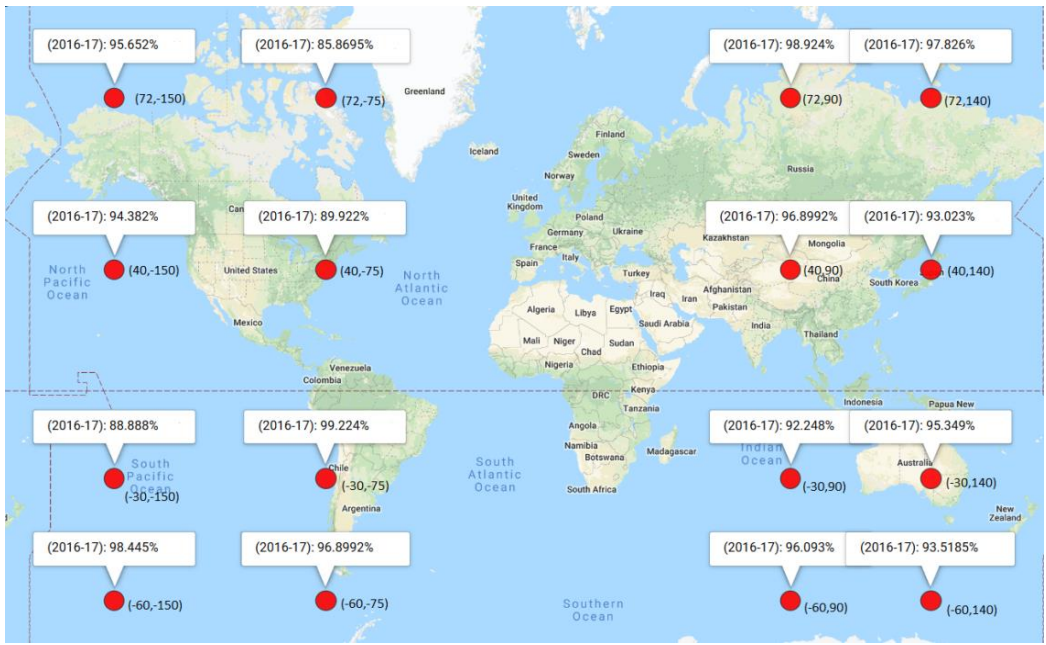

**Figure 8: Accuracy Results for 2016-2017**

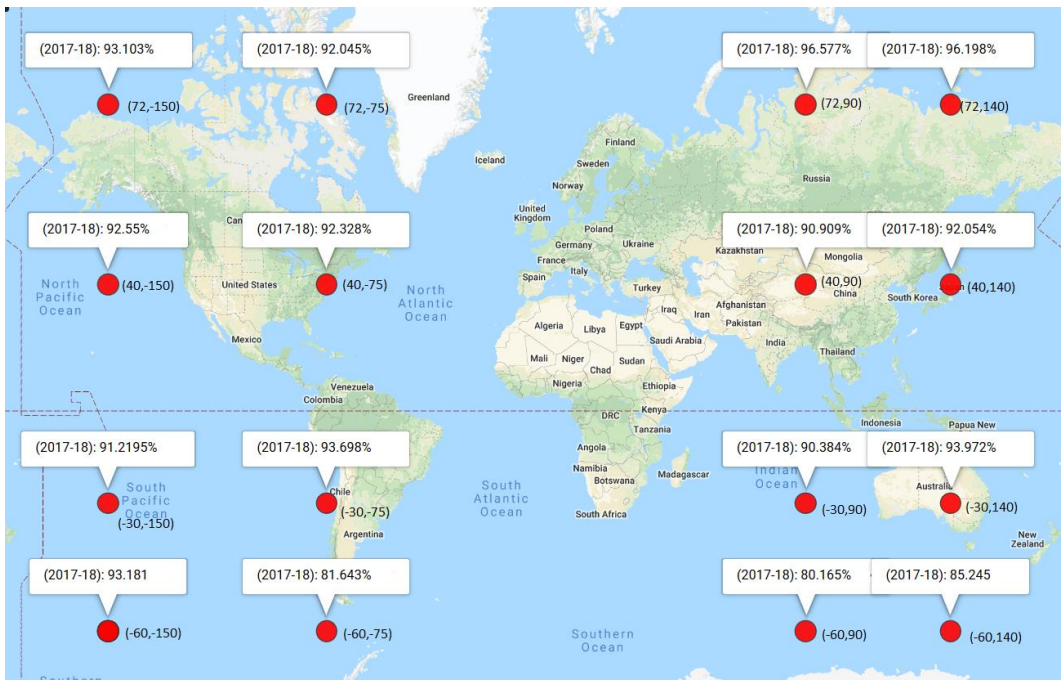

**Accuracy results for 2017-2018**

**Average Accuracy over Latitude and Longitude:**

From the 16 selected locations, we calculated the average accuracy of the classifier over given latitude (Figure 10) and longitude (Figure 11). This gives us an idea about the overall reliability of our classifier.

We had come across many research papers in which many had not considered to experiment with several other locations and maintain similar accuracy.

The variance in accuracy with latitude is mainly to due to the change in climatic conditions and weather as we move away from the equator. We plan to do further research considering the seasonal aspects and the ambience.

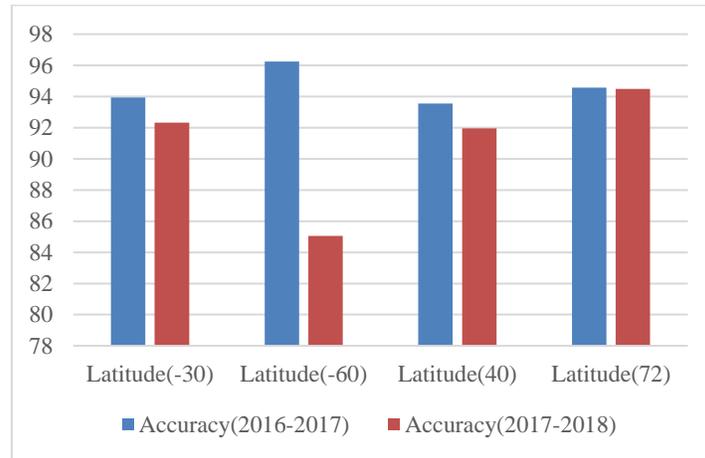

Figure 10: Average Accuracy over a Latitude

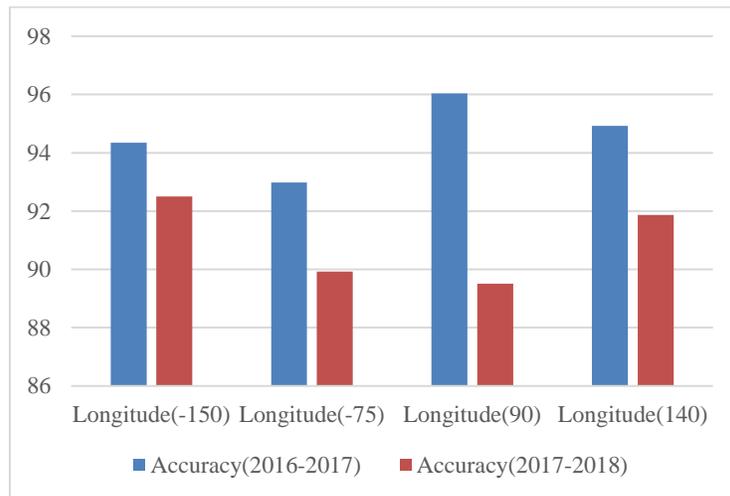

Figure 11: Average Accuracy over a Longitude

There is a noticeable drop in accuracy from 2016-2017 to 2017-2018. This can be due to unpredictable weather and ever changing climate. This could be improved in future by considering an approach involving other meteorological parameters and training the classifier with respect to seasons. Having a unique model trained for each individual season would further increase the accuracy. The accuracy also falls as we move closer towards extreme latitudes which is predominantly due to unavailability of sufficient data. The average accuracy for the 16 locations came out to be around 94.5731% for 2016-2017 and 90.9517% for 2017-2018.

## Further Discussion:

In our search for sustainable energy, solar energy has turned out to be most suitable. If we are to rely on solar energy for our critical needs, prediction of the solar energy production is indeed very important task. In our search of relevant literature, we surprisingly found that the most of the research on this topic does not address either-

I. The need for prediction to be geographically applicable to all places.

II. It was not possible to extend models to diverse locations.

III. Needed expensive customized sensor setups.

IV. Lacked sufficiently powerful context to afford superior predictive capability.

Our research overcomes all these practical hurdles by relying on NASA observatory historical data that are relevant indicators of solar energy prediction. We are certain that such extensive experimentation hasn't been carried out yet.

Since the API support begins to add additional parameters and information, we can give users the choice to select the parameters they want and build their customized dataset. The ability to pick attributes to train the classifier can also be executed once the users have a big enough collection of parameters. Creating a selection algorithm to best ascertain the variables which would have an impact on the value to be called will help in achieving the best possible accuracy.

We plan to extend our work in multiple directions and we are keen to experiment with different seasons and extended periods between training and test data. Our Machine Learning algorithms can be made more sophisticated by using more data. We currently only use spatial context in our representation – we could extend our representation to include temporal context and make it more effective by leveraging seasonal variations in the solar energy. Finally, our system can be more extensively validated from diverse locations to validate our initial results.